\documentstyle[aps,preprint]{revtex}
\draft
\tightenlines
\begin{document}

\title{Ising Meets Ornstein and Zernike, Debye and H\"uckel,
Widom and Rowlinson, and Others}
\author{Ronald Dickman$^*$}
\address{
Departamento de F\'{\i}sica, ICEx,
Universidade Federal de Minas Gerais,\\
30123-970
Belo Horizonte - MG, Brasil\\}
\date{\today}

\maketitle
\begin{abstract}
The name Ising has come to stand not only for a specific model, but 
for an entire universality class - arguably the most important such
class - in the theory of critical phenomena.  I review several examples,
both in and out of equilibrium, in which Ising universality appears
or is pertinent.  The ``Ornstein-Zernike" connection concerns a
thermodynamically self-consistent closure of the eponymous relation,
which lies at the basis of the modern theory of liquids, as applied to
the Ising lattice gas.  Debye and H\"uckel founded the statistical 
mechanics of ionic solutions, which, despite the long-range nature of 
the interaction, now appear to exhibit Ising-like criticality.
The model of Widom and Rowlinson involves only excluded-volume 
interactions between unlike species, but again belongs to the Ising
universality class.  Far-from-equilibrium models of voting behavior,
catalysis, and hysteresis provide further examples of this ubiquitous
universality class.
\end{abstract}

\pacs{$^*$email: dickman@fisica.ufmg.br}

\section{Introduction}

Since the playful title of this review might generate
confusion, let me start by saying that I have no idea whether or not
Ernst Ising actually met any of the other scientists mentioned.
The world lines intersecting here belong to models and theories,
not to persons, real or imagined!  Much as the name Galileo has
come to define a specific kind of relativity, Hamilton a kind of
dynamics, and Gauss a distribution, so
``Ising" is now indelibly associated with a specific kind of
critical behavior, a ``universality class" in the familiar jargon of
renormalization group theory.  The latter tells us (as is borne out
by experiment, analysis, and simulation), that the factors determining
scaling properties in the
neighborhood of the critical point are few, and pertain to
very general properties of the system, such as its dimensionality
and that of the order parameter, range and symmetry of interactions.
For historical reasons, the class defined by a scalar order parameter, $\phi$,
short-range interactions, isotropy, and symmetry under inversion ($\phi \to -\phi$
in the absence of an external field),
is commonly known as the Ising universality class.  While we should
expect many systems to fall in this class, most of the present article
is devoted to examples whose membership is surprising or
controversial.

My first example, self-consistent Ornstein-Zernike theory (Sec. II), is not a
bona-fide member of the Ising class, but is, in a sense, making an
impressive effort to be one!  Criticality in electrolyte solutions, discussed
in Sec. III, remains controversial, though recent theoretical and experimental
studies support Ising-like behavior.  We will see that the simplest
{\it lattice} version of the well-known primitive model may exhibit
a line of Ising-like critical points.  We are used to thinking of the Ising model
as possessing two relevant variables (or scaling fields), the temperature and
the external magnetic field.  The Widom-Rowlinson model (Sec. IV)
(in its lattice-gas version) illustrates the possibility of Ising criticality
in a system with no temperature to speak of.  
Having dispensed with temperature, we proceed in Sec. V to the more
radical step of eliminating the Hamiltonian (and thermodynamics)
altogether, in several far-from-equilibrium stochastic processes.
The guiding principle of symmetry again permits one to
understand the appearence of Ising criticality in these systems.
A brief summary is provided in Sec. VI.

\section{Self-Consistent Ornstein-Zernike Theory for the Ising Model}

Ornstein-Zernike theory \cite{oz} represents an important, early chapter in
the theory of critical phenomena, and is still widely used to 
interpret data from scattering experiments.  The Ornstein-Zernike
relation (OZR), moreover, serves as the starting point for modern
theories of liquids \cite{hanmc}.  This relation introduces the {\it direct correlation 
function} $c(r)$ \cite{note1} in terms of the total correlation function $h(r) \equiv g(r) -1$,
where $g(r)$ is the radial distribution function:

\begin{equation}
h(r) = c(r) + \rho \int d{\bf r}' c(r) h(|{\bf r} -{\bf r}'|) \;.
\label{ozr}
\end{equation}
Here $\rho$ is the number-density.  Of course one must now
supply a {\it closure}, i.e., a second relation between $c(r)$, the
intermolecular potential $w(r)$, and (in certain cases) $h(r)$ itself.
This, in general, can be done only in a heuristic manner, as for example
in the mean-spherical approximation, $c(r) = -w(r)/k_BT$, or the
more sophisticated Percus-Yevick or hypernetted chain 
closures \cite{hanmc,bal}.
Surprisingly, these simple closures lead in many cases to good
predictions for $h(r)$, and for thermodynamic properties,
via one of the standard, exact relations 
(the so-called virial, compressibility, and energy routes) involving $h(r)$.
This is remarkable, since neither the OZR nor the closure 
include any thermodynamic input.  The OZ formalism and the resulting
`integral equations' for $h(r)$ are couched purely in terms of
correlations, without reference to a free energy.  

The absence of thermodynamic input becomes evident when
one compares the predictions from different routes.  Thus the
Percus-Yevick closure yields two different equations of state
for the hard-sphere fluid, depending on whether the link to
thermodynamics is made using the virial or the compressibility
route.  Such inconsistencies suggest a new strategy for closing
the OZR, in which we use a thermodynamic relation, expressing
equality of (for example) the pressure evaluated via different routes,
to find a second relation between $h$, $c$, and $w$.
Such an approach was proposed over twenty years ago by
H$\mbox{\o}$ye and Stell \cite{hoyst}, but detailed numerical implimentations
have appeared only in the last few years.  The first such application
was, naturally, to the three-dimensional Ising model, which we
outline here.  (For details the interested reader may consult
Refs. \cite{scoza1} and \cite{scoza2}.)

We consider the Ising lattice gas with independent variables $\rho$ 
(number of particles per site) and
$\beta \equiv 1/k_BT$.  The interparticle potential $w(r)$ is zero for $r > 1$,
-1 for $r=1$ (an attractive nearest-neighbor interaction) and infinite for
$r=0$.  The latter implies the `core-condition' $h(0) = -1$, expressing the
simple fact that two particles may not occupy the same site.
On a lattice the virial route, which involves the spatial derivative of the
potential, does not exist.
Consistency between the two remaining routes, energy and compressibility,
is embodied in the relation 

\begin{equation}
\rho \frac {\partial ^2 (\rho u)}{\partial \rho^2}  = 
\frac {\partial ^2 (\beta p)}{\partial \beta \partial \rho},
\label{concond}
\end{equation}

\noindent where $u$ is the internal energy per particle and
$p$ is the pressure.  The energy relation is
\begin{equation}
u = -\frac{1}{2} q \rho (1+h_1), 
\label{en}
\end{equation}
where $h_1$ denotes the total correlation function at the nearest-neighbor
separation, and $q$ is the coordination number.  
The inverse compressibility is related to $c$ via
\begin{equation}
\frac{\partial (\beta p)}{\partial \rho} = 1 - \rho \tilde{c}(0) 
\label{invcmp}
\end{equation}
where $\tilde{c}$ denotes the Fourier transform.
The direct correlation function is, in effect, defined by the OZR,
which reads, on lattice,

\begin{equation}
h_{\bf i} = c_{\bf i} + \rho \sum_{{\bf j} \in \cal{L}} c_{{\bf i-j}} h_{\bf j} \; ,
\label{oze}
\end{equation}
(the sum is over the sites of lattice $\cal{L}$).  This relation, or, more specifically,
its Fourier representation,
\begin{equation}
1 + \rho \tilde{h} = \frac {1}{1 \! - \! \rho \tilde{c}} \;\; ,
\label{ozef}
\end{equation}
enables us to find $h$ given $c$.  Since we already have $h(0) = -1$,
the above relations, incorporating thermodynamic consistency,
suffice for determining one free parameter, as a function of
$\rho$ and $\beta$.  At this stage we introduce the {\it unique}
approximation of our theory, which is to set $c(r) \equiv 0$ for
$r > 1$, i.e., beyond the range of the interaction.  
This defines what we call the {\it self-consistent Ornstein-Zernike
approximation} or SCOZA.
After some
manipulations we obtain a nonlinear partial
differential equation for $c(1) = c_1(\beta, \rho)$
($c(0)$ is effectively fixed by the core condition).
Integrating this PDE numerically eventually yields
$h(r)$ and all thermodynamic properties of the lattice gas.

The results compare remarkably well against the best numerical
(series expansion) estimates for the lattice gas \cite{scoza1,scoza2}.  
We find the critical temperature from
the condition of a diverging compressibility at $\rho = 1/2$
(below $T_c$, infinite compressibility signals the spinodal);
the coexistence curve is constructed in the usual manner.
Especially impressive are the predictions for critical
parameters, listed in Table I; SCOZA reproduces the best
series estimates to within 0.2\%.

 \begin{table} [h]
 \begin{center}
  \caption{Comparison of critical parameters}
  \label{tab1}
  \begin{tabular}{|l|l|l|l|l|l|l|l|l|}
lattice  &  $\beta_{c, scoz} $ & $\beta_{c, be}  $ & $u_{c,scoz}$ & 
$ u_{c, be} $
& $s_{c, scoz}$ & $s_{c,be}$ & $(\beta p)_c$ & $(\beta p)_{c,be}$ \\ \hline
{\small SC}  & 0.88497 & 0.88662$^a$ & -2.010806 & -1.9961$^b$ &
1.1037 & 1.1158$^b$ & 0.1140 & 0.1124 \\
{\small BCC} & 0.62848  & 0.62947$^c$  & -2.564460 & -2.5464$^b$  & 
1.1542 & 1.1641$^b$ & 0.1256 & 0.1244 \\
{\small FCC} & 0.40772 & 0.40825$^c$   & -3.768955 & -3.7423$^b$   &
1.1703 & 1.1804$^b$ & 0.1302 & 0.1292 \\
  \end{tabular}
 \end{center}
$^a$ Ref [8]
$^b$ Ref [9]
$^c$ Ref [10]
\end{table}

Can SCOZA predict critical exponents?  Its thermodynamic
predictions are so accurate that {\it effective} exponents
(i.e., derivatives of the compressibility, order parameter, etc., with respect to
temperature)
are in reasonable agreement with Ising model values near, 
but not asymptotically close to, the critical point.
We really should not expect SCOZA to reproduce Ising critical
exponents: $c(r)$ is not zero for $r>1$ in the Ising model.
In particular, it develops a power-law tail at the critical point.
The actual critical behavior of the SCOZA-lattice gas is
unusual \cite{scoza2}: for $T > T_c$ the exponents are those of
the spherical model: $\gamma = 2$, $\delta = 5$, and $\alpha = -1$
(the specific heat does not diverge at $T_c$),
as compared with $\gamma \simeq 1.24$, $\delta \simeq 4.8$, and
$\alpha \simeq 0.1$ for the three-dimensional Ising model.
{\it Below} $T_c$, SCOZA yields a new set of exponents:
$\gamma' = 7/5$, $\beta' = 7/20$ and $\alpha' = -1/10$,
considerably better than mean-field or spherical model values.
 
Clearly the way to improve SCOZA is to relax the truncation of
$c(r)$.  But even with its current limitations, SCOZA represents the
first theory to yield globally accurate thermodynamic and
structural properties for a fluid model.  It has been applied with
success to off-lattice fluids \cite{wild}, and fluids in a disordered
matrix \cite{rosen}, and promisses to become an important tool
in the study of liquids.

\section{Ising Universality and the Primitive Model of Electrolytes}

The critical behavior of electrolyte solutions remains a
challenging area, both experimentally and theoretically.
Experiments on various ionic solutions have yielded either
mean-field like critical exponents \cite{singh}, Ising-like
exponents \cite{japas}, or a crossover from mean-field to
Ising-like behavior as one nears the critical temperature \cite{narayan}.
In some cases the crossover from mean-field to Ising exponents 
occurs at a reduced temperature much smaller than that
typically seen in non-Coulomb liquids.

Theoretical studies have tended to focus on the simple
system introduced by Debye and H\"uckel in 1923 \cite{dh},
commonly known as the {\it restricted primitive model} (RPM).
Here the positive and negative ions are represented by hard spheres
of diameter $\sigma$, containing point charges $\pm q$ at their
centers.  The solvent is not considered explicitly; its effect is
represented solely by the dielectric constant $\epsilon$ that enters the
expression for the electrostatic potential.  While clearly a minimal
model for ionic solutions, the RPM already presents great
difficulty to theory and simulation.  Debye and H\"uckel's analysis
of the dilute, high-temperature regime led to the fundamental
result that electrostatic interactions are {\it screened} on a length
scale $\Gamma^{-1}$, where $\Gamma^2 = 4 \pi \rho \beta q^2/\epsilon$.
(The effective interaction or potential of mean force 
takes on a Yukawa form $\propto e^{-r/\Gamma}/r$.)

What sort of phase diagram should we expect for the RPM?
The hard-sphere part of the potential alone will lead to a fluid-solid
transition.  At high temperatures ($T^* \equiv \sigma k_BT \epsilon /q^2 \gg 1$)
we expect packing considerations to dominate, so that the high-density
structure is FCC or HCP.  But since these structures are incompatible
with antiferromagnetic order (i.e., positive and negative charges 
occupying distinct sublattices), we should expect a structural phase transition
to a bipartite lattice (presumably BCC) as we lower the temperature,
at high density.

The opposite corner of the $\rho -T$ plane, low temperature and density
much less than the solid, is where we find the critical point
in simple fluids.  Since the RPM lacks the short-range
attraction that drives the liquid-gas transition, it is not immediately
obvious that it should exhibit such a transition.  Nevertheless,
Stell, Wu, and Larsen, using liquid-state theory, reached the
conclusion that the RPM has a liquid-gas coexistence curve 
with a critical point \cite{swl76}, but with a
critical density much lower than for a simple argon-like fluid.  
Monte Carlo simulations confirm these conclusions, but only in the last
few years have the studies of different groups
converged toward common values for the critical density and 
temperature of the RPM \cite{caillol,valleau,orkoulas}.
Given the difficulty in simply locating the critical point, it is not
surprising that current simulations shed little light on its nature.

On the theoretical side, however, there has been considerable discussion of 
RPM criticality, pointing toward Ising-like 
behavior \cite{haf,stell92,fisher94}.  This seems at odds with
conventional wisdom (Ising universality for short-range, and
mean-field behavior for long-range interactions), but may be
understood intuitively as follows.  At the low temperatures of interest
(note that $T^*_c \approx 0.05$), the Coulombic interaction strongly
suppresses charge-density fluctuations on scales larger than $\sigma$;
ions associate into pairs and larger aggregates.  Thus the effective
interacting units are not individual ions but clusters that are typically
neutral (or nearly so), interacting via multipolar forces (presumably
quadrupole and higher) that are of short range and, on average,
attractive (since such fluctuations lower the energy).

Such an intuitive picture finds support in recent experiments showing
Ising critical exponents at an ionic-solution critical point \cite{narayan}.
What is needed, from the theoretical standpoint, is an argument that
takes us from the RPM to a continuum
description of density fluctuations, $\delta \rho({\bf r})$, having the same
form (up to irrelevant terms) as that for the Ising model, i.e., the
usual $\phi^4$ field theory.  It is clear, on the other hand, that the RPM
needs to be described in terms of {\it a pair} of coupled fields, the
mass or number density $\delta \rho$ and the charge density $\psi$.
An important first step is the proper formulation of a mean-field 
theory \cite{levin,ciach}.  Very recently, Ciach and Stell constructed
a Landau-Ginzburg free energy functional, starting from the mean-field
theory of the RPM, in terms of the two fields, $\delta \rho$ and $\psi$.
They find that integrating out the charge density fluctuations leads to
an effective field theory for $\delta \rho$ having the expected $\phi^4$
form.  (Essentially, an attractive effective interaction between mass-density
fluctuations is mediated by charge-charge correlations.)  Thus the
crucial link between the RPM and Ising-like criticality appears to be at hand.

In light of the great difficulty of RPM simulations, it seems useful to
study a {\it lattice} restricted primitive model (LRPM) \cite{ds99,panag99},
since lattice simulations offer substantial economies in the computation time
required for evaluating the potential.  (Overlap checks are trivial
using a site-occupancy matrix, while the Coulomb potential - with
the contributions from the infinite periodic array of cells suitably
accounted for - may be stored in a lookup table.)
Dickman and Stell \cite{ds99} considered a lattice gas of particles interacting via site 
exclusion (multiple occupancy forbidden) and a Coulomb interaction 
$u(r_{ij}) = s_i s_j /r_{ij}$, where $r_{ij} = |{\bf r}_i - {\bf r}_j|$ is the 
distance separating the particles
(located at lattice sites ${\bf r}_i $ and ${\bf r}_j $), 
and $s_i = +1$ or $-1$ is the charge of particle $i$.   (Exactly 
half the particles are positively charged, half
negative.  They are restricted to a simple cubic lattice with periodic boundaries.
)

At full occupancy, the LRPM may be viewed as an antiferromagnetic Ising model
with long-range interactions.  One naturally expects a critical or N\'eel point
separating a high-temperature phase from a phase exhibiting antiferromagnetic
order.  (The Ising analogy facilitates formulation of a mean-field theory
for the LRPM.)
Indeed, it was proven some time ago that the LRPM on the simple cubic lattice
exhibits long-range order at sufficiently low temperatures and high 
fugacities \cite{lieb80}.
In simulations, we find that this point marks one terminus of a line
of critical points in the $\rho - T$ plane (see Fig. 1).  
The other end of the critical line is a
{\it tricritical} point, which intersects the coexistence curve between a
low-density disordered phase and the ordered phase.
The Monte Carlo simulations used a relatively small lattice (16$^3$ sites),
sufficient to map out the phase diagram by studying the order parameter,
specific heat, and correlation functions, but too small to yield
information on critical behavior \cite{ds99}.

Thus the simplest version of the LRPM shows a phase diagram that is
rather different than the continuous-space RPM, exhibiting a critical line
and tricritical point rather than a simple critical point.
One reason for this difference is that the simple cubic lattice
facilitates antiferromagnetic order.  It may be
that on a lattice that frustrates such order (the FCC structure, for example),
the critical line will disappear, and the tricritical point become a
critical point.  On the other hand, Panagiotopoulos and Kumar found
exactly this, on the simple cubic lattice, when the hard-core
exclusion range is $\geq 3$ times the lattice spacing \cite{panag99}.
A suitably modified version of the LRPM may therefore be useful 
for studying ionic criticality.  The phase diagram appears to be highly sensitive to
changes in lattice structure or short-range interactions, as was also
noted by Ciach and Stell \cite{ciach}.

Even if the critical line of the simple-cubic LRPM is not exactly
what we had in mind for understanding the off-lattice RPM, its nature, and that of
the associated tricritical point, are of interest, and potentially relevant to
other systems with Coulombic interactions.  Given the symmetry of the 
system, the Ising universality class again seems the natural candidate.
The (very limited) simulation data seem to indicate $\beta \leq 0.326$,
(the 3-d Ising value), which should in any case rule out a mean-field
type transition.  Clearly larger-scale simulations and finite-size scaling
analysis are in order; a low-temperature expansion for the fully
occupied case might also prove useful.  We close this section with
the observation that Ising criticality is compatible with a long-range
bare interaction, provided the effective interactions between critical fluctuations 
are of short range.

\section{Ising Without Temperature: The Widom-Rowlinson Model}

In this section we again consider a model imported to the lattice
from its original continuous-space formulation.
The Widom-Rowlinson (WR) hard-sphere mixture
is perhaps the simplest binary fluid model
showing a continuous unmixing transition, and has been the subject of
considerable study regarding its thermal and  interfacial properties 
\cite{wr70,wrbk}, as has the Gaussian $f$-function version of the model
introduced somewhat earlier by Helfand and Stillinger \cite{hlst}.
Despite this interest, however, definitive results on the location and nature
of the WR critical point are lacking. 
As a first step in this direction, Dickman and Stell performed
extensive simulations of the lattice-gas analog of the WR hard-sphere
mixture the
Widom-Rowlinson lattice model (WRL) \cite{ds95}.

In the original WR model, AB pairs interact {\em via} a hard-sphere
potential whilst AA and BB pairs are noninteracting \cite{wr70}.
The WRL model is a two-component lattice gas in
which sites may be at most singly occupied, and in which nearest-neighbor A-B
pairs are forbidden.  Like the WR model, this is evidently an athermal model 
(all allowed configurations 
are of the same energy), and is characterized solely by the densities of the two
species or by the corresponding chemical potentials 
$\mu_A$ and $\mu_B$.  ($\mu_A = \mu_B = \mu_c$ of course, at the critical
point.)

Despite the absence of a temperature or energy scale, the WRL model
has a close affinity to the Ising model.  To see this, note that
the WRL may be viewed as an extreme member of a family of {\em binary
alloy models} with nearest-neighbor
interactions that are repulsive between unlike species 
(interaction energies $\epsilon_{AB} > 0$,
$\epsilon_{AA} = \epsilon_{BB} = 0$). 
The Ising model may be transcribed into such a
model by identifying up and down spins with A and B particles, 
respectively, yielding a ``close-packed" alloy that unmixes at the 
Ising critical temperature.  Allowing a small fraction of vacant sites results 
in a dilute binary alloy (DBA) with a 
somewhat depressed critical temperature; continuing the dilution process, one
arrives at a model with $T_c = 0$.  This zero-temperature terminus of the DBA
critical line is precisely the WRL critical point.  
The coexistence surface of the binary alloy model in the $\rho_A$,
$\rho_B$, $T$ solid is shown schematically in Fig. 2.
One is then led to ask whether
the entire critical line shares a common behavior, or whether its character
changes at some point.  Although the former is clearly favored
on the basis of universality, a careful examination of this question 
nevertheless appears worthwhile.  

As shown in Fig. 2, the Ising coexistence curve is the intersection
of the critical surface with the plane $\rho \equiv \rho_A +\rho_B = 1$.
For $\rho < 1$ we can cross the critical line in the temperature or
the density direction, maintaining all the while $h \equiv \mu_A - \mu_B = 0$
($\mu_i $ is the chemical potential of species $i$).
Thus, while in the WRL there is no 
temperature {\em per se}, $\mu \equiv \mu_A + \mu_B$  is a
temperature-like variable, so that along the symmetry line $h=0$,
and in the vicinity of the critical point $\mu_c$,
the susceptibility should scale as $\chi \simeq (\mu - \mu_c)^{-\gamma} $, the order
parameter, $\rho_A - \rho_B \simeq (\mu - \mu_c)^{\beta} $, (for $\mu > \mu_c)$, 
and so on \cite{hlst}.  
(Note that if the chemical potentials are taken as independent variables,
no ``Fisher renormalization" of critical exponents is expected, as it would
be if we worked with fixed densities \cite{fren}.)

Ref. \cite{ds95} reports Monte Carlo simulations of the WRL in the
grand canonical ensemble, using lattices of up to 160$^2$ sites in
two dimensions and $64^3$ sites in three dimensions.  The algorithm
employs three kinds of moves:
``flips" (change of the state, A, B, or vacant, at a single site),
``exchanges" (between any pair of sites in the system), and flips 
($A \rightleftharpoons B$)
of entire clusters (note that this is always possible in the WRL since
a cluster of occupied sites is always surrounded by a border of
vacant sites).  The results for the reduced fourth cumulant, order
parameter, and susceptibility are all consistent with Ising-like
behavior \cite{ds95}.  

Following the WRL study, a series analysis of a closely
related continuum model, with Gaussian Mayer $f$-functions,
(i.e., the model introduced in Ref. \cite{hlst}), was reported by
Lai and Fisher \cite{lai95}.  This study again provides good evidence
for consistency with the Ising universality class at the critical
point marking phase separation.  More recently, the generalization
of the WRL to $q$ states, with infinite repulsion between
unlike nearest-neighbor pairs has been found to exhibit critical
behavior consistent with the $q$-state Potts model, as would be 
expected from symmetry considerations \cite{wrpotts}.

The appearance of Ising-like criticality in athermal
models in fact goes back to studies of lattice gases with
nearest-neighbor exclusion (NNE), done in the mid-1960's 
\cite{gauntmf,gaunt,runnels}.  In this case there is only a single
species of particle, and the only interaction is a hard-core
repulsion assigning infinite energy to pairs of particles with
separations $\leq 1$, in units of the lattice spacing.
On a bipartite lattice in two or more dimensions, there is a
critical density above which the particles begin to occupy one of
the two sublattices preferentially,
signalling a continuous transition to a state with antiferromagnetic
order.  (The NNE lattice gas is in fact closely related to the
zero-temperature line of the Ising {\it antiferromagnet}.)
Gaunt and Fisher analyzed series expansions for the NNE lattice gas
on various two- and three-dimensional lattices, and found
$\beta \approx 1/8$, suggesting, once again, Ising 
universality \cite{gauntmf,gaunt}.
It would be worthwhile applying modern series and simulation
methods to the NNE models, to obtain more precise results on their
critical behavior.

It is perhaps worth observing that in all of the models mentioned in
the present section (and, indeed, in the hard-sphere fluid), phase
separation is driven exclusively by {\it entropy maximization}.
Thus we have the apparently paradoxical conclusion that for
large values of the chemical potential, the ordered phase has
an entropy greater than or equal to that of the disordered phase.
This shows that the naive identification of entropy with
`disorder' is not always appropiate; interpreting increased
entropy as greater freedom seems more apt.

\section{Ising Without Equilibrium: Voters, Catalysis, and Hysteresis}

The present section concerns far-from-equilibrium models, defined by
a Markovian {\it dynamics} rather than a Hamiltonian.  Since the
transition rates do not satisfy detailed balance with respect to any
reasonable energy function 
(i.e., bounded below, and with a finite number of many-body terms),
these models have no thermodynamic interpretation.  Such systems are
nevertheless widely studied using the tools of statistical physics,
as models of, for example, populations, traffic, catalysis,
and ``self-organized" criticality \cite{bjpne}.
Many of these systems exhibit transitions between a fluctuation-free
absorbing state (admitting no escape) and an active 
phase \cite{rev1,rev2,rev3}.  Models possessing
an absorbing state, and, associated with this, a non-negative
order-parameter density, do not exhibit
Ising symmetry.  Instead, they fall generically in the universality class of
{\it directed percolation}, which plays a role analogous to the Ising class 
for absorbing-state phase transitions \cite{kinzel,liggett}.

The models to be discussed in this section do, however, observe
Ising symmetry, and belong to the Ising universality class.
The fact that the Ising class cuts across the equilibrium/nonequilibrium
boundary serves to illustrate that phase transitions in far-from-equilibrium
systems are just as worthy of the name as their equilibrium
counterparts, and do not, as is sometimes asserted, represent merely an
``analogy" to ``real" (i.e., equilibrium) phase transitions.  Of course, one's
outlook will depend on whether one adopts a thermodynamic or a mathematical
definition of a phase transition.  My point is that experience with percolation,
nonequilibrium models, and deterministic as well as stochastic cellular automata
make it natural to regard, quite generally, any singular dependence of
the properties of a system of many interacting units upon its
control parameters as a phase transition.  
From this vantage, critical phenomena represent a particular class of
singularities (attended, for example, by a diverging correlation length
and relaxation time), with equilibrium critical points representing a
particular, and not especially privileged, subset thereof.

\subsection{The Majority-Vote Model}

The prototypical example of a nonequilibrium stochastic process
with local interactions, isotropy, and up-down symmetry is the 
majority-vote model \cite{liggett}.
At each site of a lattice there is a ``spin" variable $\sigma_i = \pm 1$.
A Markov process is defined as follows.  At each step, a site $i$ is chosen
at random, and $\sigma_i \to \sigma_i'$, which is taken to be equal to the
majority of its nearest neighbors (i.e., sgn[$\sum_{j nn i} \sigma_j$]) with
probability $p$, and to the minority with probability $q=1-p$.
(If there is no majority $\sigma_i$ flips with probability 1/2.)
Although the transition rates
do not satisfy detailed balance, it is not hard to see
that the parameter $q$ plays a role analogous temperature in the
equilibrium kinetic Ising model.  (The dynamics of the majority
vote model is akin to Glauber dynamics; there are no conserved
quantities.)  For $q=0$, ``voters" have no independene of opinion whatever, 
and slavishly follow the local majority.   This $q=0$ limit defines the
{\it voter model}, which has two absorbing states, all +1 or all -1,
just as in the Ising model at $T = 0$.
(The voter model belongs to a universality class
different than Ising, the so-called compact directed percolation 
class \cite{essam89,rdtre}.)  For $q > 0$ but small, the stationary state
exhibits spontaneous ``magnetization": $m \equiv \langle \sigma_i \rangle
= \pm m_0 $, with $ m_0 > 0$.  Above a critical $q$ value
($q_c \simeq 0.075$ for the square lattice \cite{mario}), the stationary state
is disordered ($m = 0$).  (No ordered state is found in one dimension,
just as for the equilibrium Ising model.)

Thus the phase diagram of the majority-vote model is qualitatively the same as
for the standard Ising model.  Confirmation that the critical behavior is
also of the Ising kind comes from Monte Carlo simulations reported by
de Oliveira \cite{mario}, which showed (for the two-dimensional case),
that the exponents $\nu $, $\gamma$
and $\beta$ are the same as those of the Ising model.
Many nonequilibrium models have now been shown to exhibit
Ising-like critical behavior, indicating that this universality class is equally
robust, in or out of equilibrium.  Examples include a family of
generalized majority-vote models \cite{oms}, an anisotropic majority-vote
model \cite{teixeira}, a two-state immunological model \cite{tome96}, 
and various two-temperature Ising models \cite{garrido87}.
In the latter example, a spin system evolves via two dynamical processes,
for example, single-spin flips and nearest-neighbor exchanges,
{\it each having its own temperature} (see Ch. 8 of Ref. \cite{rev3} for a review).
Ising-like {\it short-time} critical behavior has also been 
established for certain nonequilibrium models with up-down symmetry \cite{tome98}.

The principle underlying universality of critical behavior in these examples
appears, once again, to be symmetry.  While detailed renormalization group
analyses are lacking, Grinstein et al. have argued that
models obeying up-down symmetry and exhibiting a continuous 
phase transition, be they in equilibrium or not, 
will have a coarse-grained descrption of the same 
form as for a standard kinetic Ising model \cite{gjh}.
We can see this as follows: if such a description exists, then in a
Langevin-like equation for the order parameter density $m({\bf r},t)$,
only odd powers of $m$ may appear (up-down symmetry).
Gradient terms are ruled out by isotropy, so the lowest-order
derivative term will be $ \nabla^2 m$.  Generically,
the transition to an ordered state will be controlled by the coefficient $a$
of the term $\propto m$.  (In mean-field theory this coefficient vanishes 
at the critical point, but in a full analysis the nonlinear terms will 
renormalize $a_c$ to a nonzero value.)
That is, the order-parameter density will obey (in the nonconserving case)
a time-dependent Landau-Ginzburg equation of the form

\begin{equation}
\frac{\partial m({\bf r},t)}{\partial t} =  \nabla^2 m -a m - b m^3 + \eta({\bf r},t) \;,
\label{lge}
\end{equation}
where we have dropped higher-order terms ($\propto m^5$, etc.) since they are
irrelevant to critical behavior.  The noise term $\eta({\bf r},t)$ is zero-mean, and
Gaussian, with autocorrelation 
$\langle \eta({\bf r},t) \eta({\bf r}',t') = \Gamma \delta^d ({\bf r} - {\bf r}') \delta(t-t')$.
In equilibrium $\Gamma$ is of course proportional to temperature.  Out of
equilibrium no such relation exists, but the theory again predicts an Ising-like
transition; the deviation from equilibrium is found to be irrelevant, near the
upper critical dimension \cite{gjh}.
The principle that systems with the same symmetries
share a common critical behavior, whether in or out of equilibrium,
has recently been extended to models with Potts-like 
symmetries \cite{brunstein,tome00}.

We have seen that a variety of nonequilibrium perturbations preserve the 
character of the transition in spin systems or lattice gases.  It is worth noting
that biasing the hopping rates to favor movement along a particular axis
(e.g., in a lattice gas with attractive interactions and a Kawasaki-type
exchange dynamics) breaks an essential symmetry of the Ising model by
introducing a preferred direction.  Such ``driven diffusive systems", as they have 
come to be called, exhibit critical behavior outside the Ising class.  The precise
nature of the transition remains controversial; for reviews see Refs. \cite{rev3}
and \cite{schmitt}.

\subsection{Catalysis Models}

Another large class of nonequilibrium lattice models arises in the study of
heterogeneous catalysis, typically on a metallic surface.  One of the first
such models to be studied in detail was introduced by Ziff, Gulari and Barshad (ZGB),
to describe the reaction CO + 1/2 O$_2$ $\to$ CO on a platinum surface \cite{zgb}.
These models typically exhibit transitons to an absorbing state, which (when
continuous) fall
in the directed percolation class \cite{glw}; 
they are reviewed in chapter 5 of Ref. \cite{rev3}.

The ZGB model exhibits two phase transitions, one continuous, the other
discontinuous.  Under a suitable perturbation, however, the latter can become a
critical point, which appears to belong to the Ising class.
In the ZGB model, the
catalytic surface is represented by a two dimensional lattice. 
The transition rates involve a single parameter, $Y$:
the probability that a molecule arriving
at the surface is CO.
This molecule needs only a single vacant site to adsorb,
as indicated by experimental studies; O$_{2}$ requires a pair of vacant sites.
After each adsorption event, the surface is immediately cleared of
any CO-O nearest-neighbor (NN) pairs, making the
reaction rate for CO$_{2}$ formation, in effect, infinite.
The adsorption and
reaction events comprising the dynamics go as follows.
First choose the adsorbing species --- CO with
probability $Y$, O$_{2}$ with probability $1-Y$ --- and 
a lattice site {\bf x} (or, in
the case of O$_{2}$, a NN pair, ({\bf x},{\bf y})), at
random. If {\bf x} is occupied (for O$_2$, if {\bf x} and/or {\bf y} are occupied), 
the adsorption attempt fails. If the newly-adsorbed molecule is CO,
determine the set $ \cal O $({\bf x}) of NNs of {\bf x}
harboring an O atom.  If this set is empty, the newly-arrived CO 
remains at {\bf x};
otherwise it reacts, vacating {\bf x} and one of the sites in $ \cal O $({\bf x}),
(chosen at random if $ \cal O $({\bf x}) contains more than one site).
If the newly-adsorbed molecule is O$_2$, construct $\cal C$({\bf x}), the
set of neighbors of {\bf x} harboring CO, and similarly $\cal C$({\bf y}).
The O atom at {\bf x} remains (reacts) if  $\cal C$({\bf x}) is empty (nonempty),
and similarly for the atom at {\bf y}.

At a threshold value of $Y$, $ y_{2}$ ($\simeq  0.5256$ on the square lattice),
the ZGB model is said to exhibit ``CO-poisoning."  That is, the stationary
state changes abruptly from an active one ($Y < y_2$),
with continuous production of CO$_2$,
to an absorbing state with all sites occupied by CO.  (With all sites blocked,
there is no way for O$_2$ to adsorb.)  The transition is sharply discontinuous.
Experiments on surface-catalyzed CO + 1/2 O$_2$ $\to$ CO reactions do
show a discontinuous transition between states of high and low reactivity
as the partial pressure $p_{CO}$ (analogous to $Y$ in the model),
 is increased \cite{ehsasi}.  With increasing
temperature, the transition softens, and at a certain temperature becomes continuous, 
about which the reaction rate becomes a smooth function of $p_{CO}$.  
This appears to be the result of thermally-activated, nonreactive 
desorption of CO.

A similar sequence of alterations is observed if we include nonreactive desorption
of CO at a certain rate, $k$, in the ZGB model.  
With CO desorption
there is no longer a CO-poisoned state, but for small $k$ the
discontinuous transition between low and high CO coverages persists. Above a critical
desorption rate $k_c \simeq 0.0406$, the coverages (and the CO$_2$ production rate)
vary smoothly with 
$Y$ \cite{bros,evans92,tome93}.  Simulation results
indicate that at the critical point of the CO transition,
the critical exponent $\nu = 1$, and the reduced fourth cumulant $u \simeq 0.61$,
as would be expected for the two-dimensional Ising model \cite{tome93}.

\subsection{Hysteresis in the Ising Model}

For our final example we come full circle to the Ising model, 
forced out of equilibrium, this time, not through a bias or conflicting dynamics,
but by a {\it time-dependent external field} $h(t)$.  Of principal interest is
a periodically varying field, for example, $h(t)$ sinusoidal or a square wave,
with 
\begin{equation}
\langle h \rangle \equiv \frac{1}{{\cal T}} \int_0^{\cal T} h(t) dt = 0 \;,
\label{havg}
\end{equation}
where ${\cal T}$ is the period of oscillation.
Well below the critical temperature, and at sufficiently low frequencies,
we expect to observe a hysteresis loop in the magnetization-field plane.
As we raise the frequency, the system will at some point
be unable to follow the rapid variations of the field, and 
the magnetization $m(t)$ will remain close to one of its stationary values,
i.e., $+m_0$ or $-m_0$.  That is, for high frequencies we expect

\begin{equation}
Q \equiv \frac{1}{{\cal T}} \int_0^{\cal T} m(t) dt \neq 0 \;
\end{equation}
whereas at low frequencies $Q$ should be zero.

For a sinusoidal field, the time-dependent mean-field equation 
describing this system predicts a continuous transition between
a dynamically disordered ($Q=0$) and a dynamically ordered ($Q \neq 0$)
state at a critical frequency, which depends on the temperature
and $h_0 = \max[h(t)]$ \cite{tome90}.  The transition was confirmed
in various Monte Carlo simulations, and may be relevant to experiments
on thin magnetic films \cite{acharyya,chakrabarti}.

Rikvold and coworkers suggested that the transition between
dynamically ordered and disordered states occurs when the period
${\cal T}$ becomes comparable to the metastable lifetime $\tau (h_0,T)$
of the state with the ``wrong" sign of the magnetization (in a static field of
magnitude $h_0$), and
presented extensive simulation results to support this proposal \cite{rikvold}.
These authors made a detailed numerical study of the critical behavior
of $Q$ in two dimensions; their results for exponent ratios and the
fourth cumulant are consistent with Ising values. While
the very general symmetry arguments outlined above again favor
such a conclusion,   
finding a path from the microscopic dynamics of this nonstationary
process to an equation of the
form of Eq. (\ref{lge}) seems a particularly challenging theoretical task.

\section{Summary}

I have reviewed a number of examples in which critical behavior
in the Ising universality class appears, despite the presence of
long-range interactions, lack of a temperature scale, or lack of
equilibrium.  The Ornstein-Zernike approach that comes closest
to reproducing Ising-like behavior was also reviewed.  A general
conclusion is that symmetries - of the order parameter and
of the dynamics - are essential to determining critical behavior;
the question of equilibrium is of minimal, if any, significance.  On the
other hand, we have seen that the effective, coarse-grained dynamics
can look quite different from the microscopic interactions.  In many cases
there remains a considerable gap between these two levels of description.
\vspace{1em}

\noindent {\bf Acknowledgements}
\vspace{1em}

It is a pleasure to thank George Stell and T\^ania Tom\'e for 
collaborations that led to some of the results described in this
work.  This work was supported by CNPq, Fapemig and MCT.

\newpage
\noindent Figure Captions

\noindent Fig. 1.
Best estimates for the location of the critical line and
coexistence curve in the lattice RPM (simple cubic lattice).  
\vspace{1em}

\noindent Fig. 2. Schematic coexistence surface of the binary alloy
model.  The Ising model coexistence curve is the intersection of the
coexistence surface with the plane $\rho_A +\rho_B = 1$.  The 
WRL coexistence curve is the intersection with the plane $T=0$.
\vspace{1em}


\begin{thebibliography}{99}
\bibitem{note1}
	Note that $c(r)$ plays a role analogous to a two-point vertex function
	in field theory.
\end{thebibliography}
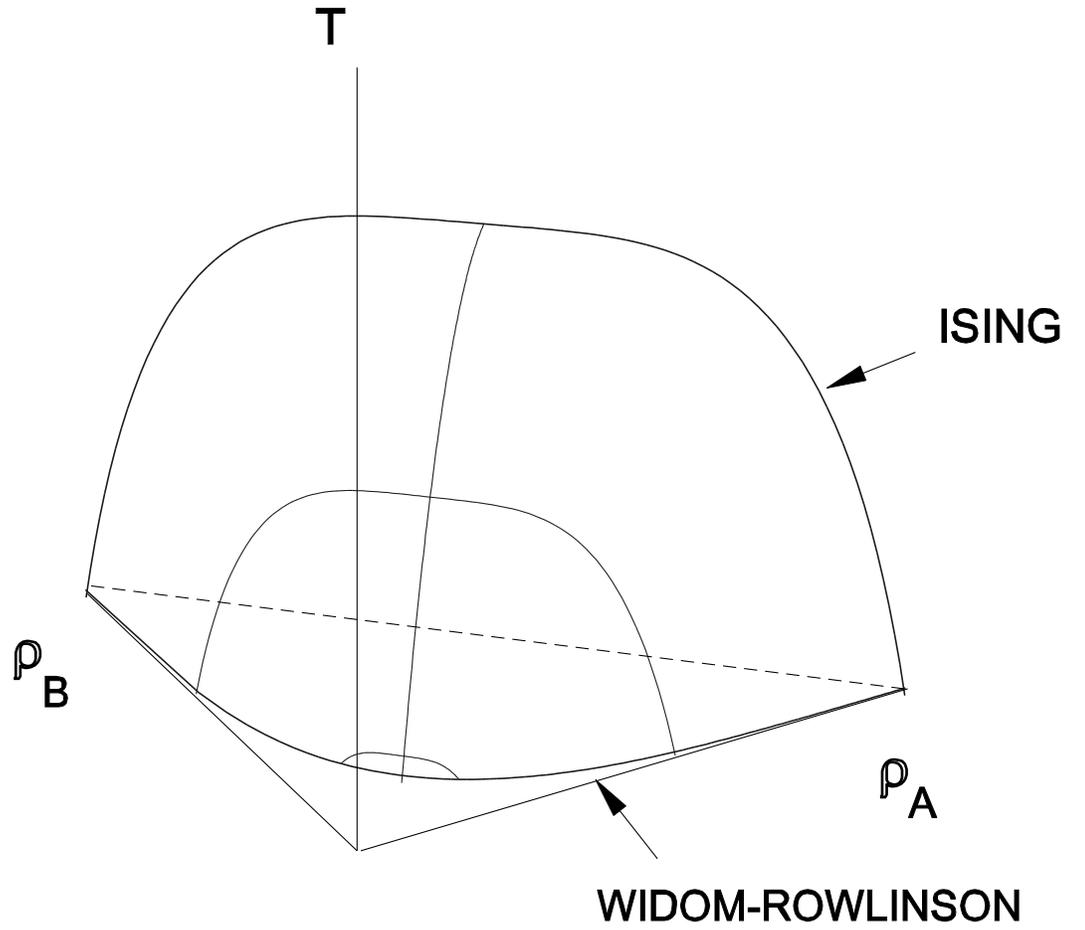
\end{document}